\documentclass[aip,reprint]{revtex4-1}
\usepackage[czech,english]{babel}
\makeatletter\AtBeginDocument{\let\@elt\relax}\makeatother
\usepackage[table]{xcolor}
\usepackage{graphicx}
\usepackage{amsmath}
\draft % marks overfull lines with a black rule on the right

\begin{document}

\preprint{}

\title{Hot Phonon Bottlenecks and the Role of Non-Equilibrium Acoustic Phonons in III-V Multi-Quantum Well Systems} %Title of paper



\author{Izak Baranowski}
\email[]{ibaranow@asu.edu}
\author{Dragica Vasileska}
%\thanks{}
\affiliation{Arizona State University}
\author{Ian R. Sellers}
\affiliation{SUNY Buffalo}
\author{Stephen M. Goodnick}
\affiliation{Arizona State University}


\date{\today}

\begin{abstract}
	The hot phonon bottleneck effect is a promising mechanism for the realization of a true hot carrier solar cell.
	Prior work has assumed that the acoustic phonons created via decay of polar longitudinal optical (LO) phonons are assumed to quickly leave the system or thermalize quickly due to multi-phonon processes.
	The present work furthers the models of previous work to include a build-up of longitudinal acoustic phonons in addition to LO  phonons due via the Klemens process.
	By including this additional process, nonphysical assumptions concerning the LO anharmonic lifetime are no longer required, resulting in a better explanation of the experimental results, and pointing towards new approaches in achieving high carrier temperatures during photo-excitation. 
\end{abstract}


\maketitle %\maketitle must follow title, authors, abstract and \pacs

\section{Introduction}
Hot phonon bottlenecks have been observed in confined III-V systems \cite{zhang_review_2022} and bulk perovskites \cite{kahmann_hot_2019,yang_acoustic-optical_2017}.
The hot phonon bottleneck is a reduction in carrier thermalization due to the build-up of a non-equilibrium optical phonon population.
Since the emission scattering rate is proportional to $N_{LO}+1$ and the absorption scattering rate $N_{LO}$, the two rates become approximately equal at high $N_{LO}$ values. 
The hot phonon bottleneck can be used for hot carrier solar cells \cite{ferry_challenges_2020}, in which the thermalization of carriers, normally assumed in the detailed balance limit \cite{shockley_detailed_1961}, is inhibited in some way; thus leading to a dramatic increase in the efficiency limit \cite{ross_efficiency_1982}.
However, the specific mechanisms behind the observed mechanisms is not well understood, and as such developing a true hot carrier solar cell has proven challenging.

Previous work has investigated this thermalization process via Ensemble Monte Carlo (EMC) methods in confined III-V systems  \cite{zou_role_2023,baranowski_monte_2025} and bulk perovskites \cite{faber_hot_2024}.
In the work presented in recent work \cite{baranowski_monte_2025}, a large optical phonon decay lifetime was needed to match the experimental results of Esmaielpour, \textit{et al.}\cite{esmaielpour_hot_2021}, due to the proper use of an experimentally calibrated ABC model for the carrier recombination rate compared with the work of Zou et al. {\cite{zou_role_2023} who used a fixed 200 ps recombination time which is artificially short, leading to additional heating.
In prior work back to the 1980s,
 the effect of acoustic phonons on the optical phonon decay was neglected with the justification that the acoustic phonons, with their relatively large group velocity, quickly left the system.
However, in the context of hot carrier solar cells, this process has not been suggested for III-V confined systems.
Prior discussions in transient simulation for both bulk and confined systems in III-V and II-VI materials suggest that a build-up of acoustic phonons can occur.   \cite{hejda_hot-electron_1993,jursenas_dense_1998,jursenas_carrier_1999,kral_electrophonon_1994,kral_hot-electron_1994,usher_theoretical_1994,zukauskas_second_1998, usher_theoretical_1994}
Dyson and Ridley\cite{dyson_role_2013} did consider a model for the build-up of LA (longitudinal acoustic) phonons under steady-state conditions and what effect that would have on the observed lifetimes of the LO phonons, however, the analysis lacks a self-consistent treatment of charge carriers in the process.

The paper is organized as follows: in section \ref{sec:model} we briefly discuss prior work and afterwards we explain our new theoretical model that takes into account non-equilibrium acoustic phonons. 
Simulation results from this study are then discussed in section \ref{sec:results}.
Important conclusions regarding the physics of the processes occurring in these structures are presented in section \ref{sec:conclusion}.

\section{Model}
\label{sec:model}
Evidence of hot phonon bottlenecks in III-V systems has been determined from time-resolved Raman \cite{grann_nonequilibrium_1996} and time-resolved photoluminescence\cite{hirst_hot_2011}.
Experimental evidence of hot carriers in confined systems can also be determined by continuous wave laser-excited photoluminescence (PL) measurements\cite{esmaielpour_enhanced_2018,sandner_hot_2023,zhang_extended_2016}.
Under laser excitation, the resulting PL is measured and fit to a modified Planck's law\cite{esmaielpour_enhanced_2018,esmaielpour_hot_2021}: 
\begin{equation}
  I_{PL}(E) = \frac{2\pi A(E)E^2}{h^3 c^2}
  \left[
 \exp\left(
 \frac{E-\Delta \mu}{k_B T}
 \right) -1
  \right]^{-1}
\end{equation}

To compare with experiment, an EMC method is used to solve the Boltzmann transport equation for photoexcited electrons and holes which approximates the conditions of laser excitation.
A constant optical generation rate is selected to give a specified steady-state well density ($D_{ss}$) for $D_{ss}$-dependent recombination model based on Shockley-Read-Hall, band to band, and Auger recombination (ABC model), calibrated to experimental density dependent measurements on similar MQW systems\cite{piyathilaka_hot-carrier_2021}.
When steady-state is reached, the distribution function, $f(E)$, of the carriers is fit to a Fermi-Dirac function across the entire distribution in order to determine the carrier temperature:
\begin{equation}
        f(E) = \frac{1}{\exp \left(\frac{E-E_f}{k_b T} \right) +1}
\end{equation}

For this work, we simulate three different Type-II AlAs$_{0.14}$Sb$_{0.86}$/InAs Multi-Quantum Well (MQW) structures with two 2.1 nm InAs wells separated by an AlAs$_{0.14}$Sb$_{0.86}$ barrier 2.1, 5.2, or 10 nanometers thick. %(see Fig. \ref{fig:struct}).
The model is general and can also be used for Type-I heterostructures as well.

\subsection{Previous Theoretical Model}
This work was based on a simulation framework that considers all relevant scattering mechanisms in this material system, including carrier-carrier scattering, degeneracy effects, and non-equilibrium phonons\cite{hathwar_nonequilibrium_2019}.
The energy eigenstates, $E$, and corresponding eigenfunctions,$\psi$, are obtained via a self-consistent solution of a Schr\"{o}dinger-Poisson equation.
In this iterative procedure, the electron and hole densities are determined from the EMC transport module.
Relevant scattering mechanisms included in our model include longitudinal polar optical (LO) phonon, transverse optical (TO) phonon, interband, intervalley ($\Gamma \leftrightarrow L$), acoustic, and carrier-carrier (e-e,e-h,h-h) scattering.
The phonons responsible for interband, intervalley, acoustic, and TO scattering are assumed to maintain their equilibrium phonon number.
These mechanisms in the optical branch (interband, intervalley, and TO) are considered as isotropic assuming parabolic dispersion.
This causes the wavevector, $\mathbf{q}$, to be uniformly distributed over the Brilloiun zone and therefore does not build up appreciably at small wavevectors as do LO phonons.
Acousic phonon scattering is treated in 2D as elastic, which also results in isotropic scattering.
The LO phonon-electron(hole) interaction, on the other hand, is a Fr\"{o}hlich interaction.
Due to the Coulombic nature of the interaction, $\mathbf{q}$ is small, and therefore confined to a small area of the Brillouin zone center.
With a sufficiently long LO phonon lifetime, $\tau_{LO}$, the LO phonons are then able to build up a substantial $N_{LO}$ and develop a hot phonon bottleneck.

In prior work\cite{lugli_nonequilibrium_1987}, non-equilibrium LO phonons are incorporated via discretizing the zone center of the LO phonon branch in $q$, initialized with the equilibrium phonon number, $N_{LO,0}$. 
When a LO phonon absorption or emission event is selected in the EMC, $N_{LO}(q)$ is decremented or increased by the ratio of the carrier and phonon density of states:
\begin{equation}
	\Delta N_{LO} = \frac{2\pi}{q^2 \Delta q S}
\end{equation}
where $S$ is the simulation area and $\Delta q$ is the $q$-space mesh spacing.
Prior work\cite{lugli_nonequilibrium_1987,lugli_monte_1989,zou_role_2023,baranowski_monte_2025} described the LO phonons decay via anharmonic processes within a relaxation time approximation:
\begin{equation}
	\frac{\partial N_{LO}(q)}{\partial t}  = - \frac{N_{LO}(q) - N_{LO,0}(q)}{\tau_{LO}} 
	\label{eq:LOdecay}
\end{equation}
Instead of re-calculating the scattering rate every time $N_{LO}$ changes, the scattering rate is calculated with a pre-selected, large $N_{LO}$ value.
A rejection step is then performed with the tabulated $N_{LO}(q)$ value to determine if the mechanism is actually selected.

The previous model is described in detail in references \cite{baranowski_monte_2025,zou_role_2023,baranowski_monte_2023}.
Specific material parameters used in this work are the same as those in Ref. \cite{baranowski_monte_2025} unless otherwise specified.

\setlength{\arrayrulewidth}{0.5mm}
\setlength{\tabcolsep}{5pt}
\renewcommand{\arraystretch}{1.5}
\begin{table}[ht]
\begin{center}
\begin{tabular}{|c|c|c|c|c|c|}
	\hline
	  &LO & TO & LA & Electrons & Holes   \\
	\hline
	LO& - & -  &\textbf{NEQ} & NEQ       & NEQ        \\
	\hline
	TO& - & -  & -  &   -       &   EQ      \\
	\hline
	LA&  \textbf{NEQ}  & -    &  -  &  EQ & EQ        \\
	\hline
	Electrons& NEQ  & -   &  EQ  &NEQ   &NEQ         \\
	\hline
	Holes    &  NEQ    & EQ   &  EQ  & NEQ  &NEQ         \\
	\hline
\end{tabular}
\end{center}
\caption{Considered interactions: bold text shows the contribution of this work to prior models}
\label{tab:interactions}
\end{table}

\begin{figure*}
\includegraphics[width=\textwidth]{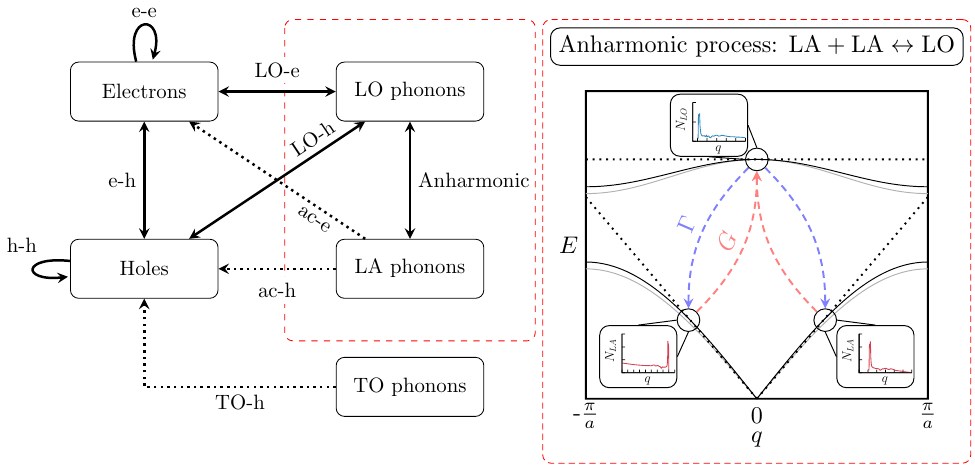}
\caption{Considered interaction used within the EMC: solid lines show interactions considered as non-equilibrium interactions, dashed lines show interactions considered as equilibrium interactions. (See  table \ref{tab:interactions} for a more detailed description). Inset shows the considered Klemens-like decay mechanism, note that the considered area of $q$-space is a small section of the Brillouin zone. The dotted lines in the inset showed the considered dispersions for LO and LA phonons.}
\label{fig:phonons}	
\end{figure*}

\subsection{Simulation Framework for Nonequilibrium Acoustic Modes}

In order to describe the $N_{LO}$ build up and decay, a Klemens decay, $\Gamma$, and generation process, $G$, is considered
\cite{ferry_decay_1974}:
	\begin{align}
	\label{eq:dnlo}
		\frac{\partial N_{LO}}{\partial t} &= -\Gamma_0 (\Gamma - G) \\
	\frac{\partial N_{ac}}{\partial t} &= \Gamma_0 (\Gamma - G) - \frac{N_{ac}-N_{ac,0}}{\tau_{ac}} 
	\label{eq:dnac}
	\end{align}
	In equations \ref{eq:dnlo} and \ref{eq:dnac}, $\Gamma_0$ is a constant decay time, which can be calculated based on physical constants depending on what mechanism is considered (\textit{e.g.} anharmonic or piezoelectric decay) \cite{ferry_decay_1974}. 
	For the purposes of comparing with prior work, $\Gamma_0$ will be set to ensure specific $\tau_{LO}$ lifetime values. 
	The $\Gamma$ and $G$ terms are then given by:
	\begin{align}
		\Gamma &= \left[N_{ac}'(q')+1\right]\left[N_{ac}''(q'')+1\right]N_{LO}(q_{LO}) \\
		     G &=N_{ac}'(q')N_{ac}''(q'')\left[N_{LO}(q_{LO})+1\right]
	\end{align}

wherein the phonon modes are selected to ensure conservation of both energy and momentum: $q_{LO} = q' + q''$ and $E_{LO} = E' + E''$.	
	A linear dispersion, $E=\hbar v_s q$ is assumed for the LA modes, and a flat dispersion, $E = \hbar \omega_{LO}$ is assumed for the LO phonons.
	To roughly approximate both diffusion of the LA phonons and their decay via multiphonon processes, a relaxation lifetime $\tau_{ac}$ is introduced.
	Acoustic phonons are relatively slow when compared to electrons and holes, \textit{e.g.}, it would take LA phonons with a group velocity of 3088 m/s on average 324 ps to leave a 1 $\mu\textrm{m}$ radius if transport was purely ballistic.
	In addition, since we are considering a confined system, the phonon transport is further inhibited.
	Simulations show that 3-phonon processes can take around 100 ps to equilibrate \cite{ono_nonequilibrium_2017}, however this could vary greatly depending on the material. 
	These observations justify the use of lifetimes that are relatively long compared to the optical phonon lifetimes.
This model does not consider broadening that would result from multiphonon acoustic processes.
Emission and absorption of acoustic phonons from carrier interaction is not considered, since electrons and holes normally interact with phonons at the center of the Brillouin zone. 
In this region, the acoustic phonons are thought to have negligible energy and most likely do not contribute to the thermalization of the carriers.
Since we are interested in general trends, more sophisticated carrier-acoustic phonon interactions are not considered.

To compare with previous work done with a constant $\tau_{LO}$, $\Gamma_0$ is set as:
\begin{equation}
	\Gamma_0(q) =  \frac{1}{ \tau_{LO} \left(1 + N_{ac,0}(q') + N_{ac,0}(q'') \right)} 
\end{equation}
which ensures that, if $N_{ac}$ were fixed to an equilibrium value,  $\frac{\partial N_{LO}}{\partial t} $ of Eq. \ref{eq:dnlo} would be equal to that of Eq. \ref{eq:LOdecay} for all $q$.
There is some $q$ dependence for optical phonon lifetimes, but the variation is fairly small \cite{bhatt_theoretical_1994}.
\section{Results}
\label{sec:results}

\begin{figure*}
	\includegraphics[width=1.0\textwidth]{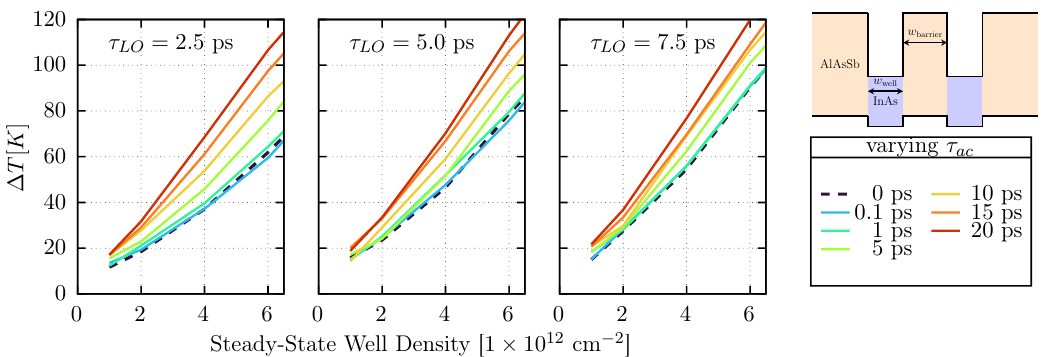}
	\caption{ Results for the $w_{\textrm{barrier}}=10$ nm structure with a constant $\tau_{LO} = 2.5$, $5.0$, and $7.5$ ps and various $\tau_{ac}$  $\Delta T$ between electron temperature and lattice temperature. }
\label{fig:tacs}
\end{figure*}

\begin{figure*}
	\includegraphics[width=1.0\textwidth]{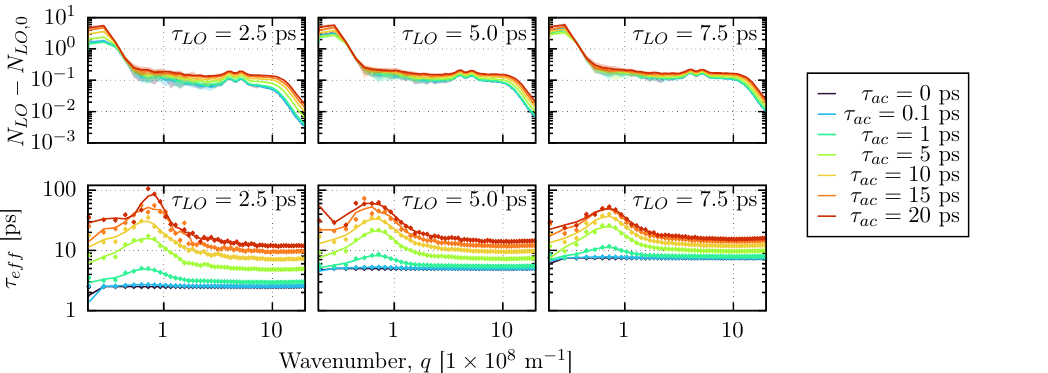}
	\caption{ Results for the $w_{\textrm{barrier}}=10$ nm structure with a constant $\tau_{LO} = 2.5$, $5.0$, and $7.5$ ps. The top row shows relative $N_{LO}$ vs $q$ at a steady-state well density of $2\times10^{12}$ cm$^{-2}$. The bottom row) $\tau_{eff}$ vs $q$ at a steady-state well density of $2\times10^{12}$ cm$^{-2}$. }
\label{fig:teffac}
\end{figure*}

To assess the impact of the non-equilibrium LA phonons on the LO phonons' lifetime, we define an effective LO phonon lifetime:
\begin{equation}
	\tau_{eff}(q) = 	\frac{N_{LO}(q) - N_{LO,0}(q)}{\Gamma_0(\Gamma-G)} 
\end{equation}
which allows us to compare the effect of adding a non-equilibrium LA phonon branch to the LO relaxation time approximation of previous works\cite{baranowski_monte_2025,zou_role_2023}.

Figure \ref{fig:tacs} shows the effect on electron temperature at steady-state with the introduction of non-equilibrium LA phonons.  
It can be seen that increases of $\tau_{ac}$ directly correspond to increases in the electron carrier temperature.
As $\tau_{LO}$ increases, $\Delta T$ is less sensitive to changes in $\tau_{ac}$.

Figure \ref{fig:teffac} shows the corresponding steady-state state phonon number, relative to the equilibrium value, and $\tau_{eff}$.
At larger $q$ values, $\tau_{eff}$ approaches the steady-state value predicted by Dyson and Ridley \cite{dyson_role_2013}.
Dyson and Ridley's model predicts that at elevated optical phonon numbers, the effective lifetime would be diminished.
However at low $q$ values, $\tau_{eff}$ unexpectedly increases.
The peak at $\approx 0.7 \times 10^8$ m$^{-1}$ does not correspond to the phonon number peak at low $q \approx 0.2 \times 10^8$ m$^{-1}$ values.
This is interesting in the context of Raman spectroscopy measurements, where the measured phonon lifetime time would correspond to the peak at $0.2 \times 10^8$ m$^{-1}$ since this is the minimum $q$ for high-energy electrons immediately after photo-excitation.

The difference in $N_{LO}$ and $\tau_{eff}$ peak positions may be due to the strong interaction with electrons at $\approx 0.7 \times 10^8$ m$^{-1}$, as this is a section of the Brillouin zone in which the LO phonons can interact with electrons over the whole energy range.
The defined $\tau_{eff}$ only considers the interactions the LO phonons and the daughter LA phonon modes via their phonon numbers.
At the $\approx 0.7 \times 10^8$ m$^{-1}$ point, the LO phonons are equilibrated between with the daughter modes and also the absorption and emission events of the electrons and holes.
The absorption and emission events are frequent enough to diminish the decay rate.

$\tau_{eff}$ is dependent on both lifetimes, $\tau_{ac}$ and $\tau_{LO}$.
The $\tau_{eff}$ shifts upward as $\tau_{ac}$ increases.
As $\tau_{LO}$ increases, the relative peak decreases and the $\tau_{eff}$ curve becomes less sensitive to increases in $\tau_{ac}$.
Since both the $G$ and $\Gamma$ process are proportional to the $\Gamma_0$ term, a long $\tau_{LO}$ then means that the generation LO phonons through the reverse process is also slower. 

Figure \ref{fig:wells} A shows the best fits for the two different well geometries ($10$ nm and $5.2$ nm AlAsSb barriers).
An improved fit compared to prior work\cite{baranowski_monte_2025} is achieved, especially for the 5.2 nm barrier case, where the ``turn-on'' behavior, referred to by Esmaielpour, \textit{et al.}\cite{esmaielpour_hot_2021} is captured.
Of particular importance is the fact that a good fit to the experimental temperature rise with injection density is achieved with a bare LO decay time of 7.5 ps and an acoustic lifetime 10 ps, rather than requiring an unphysically large single LO phonon decay time.

\begin{figure*}
	\includegraphics[width=1.0\textwidth]{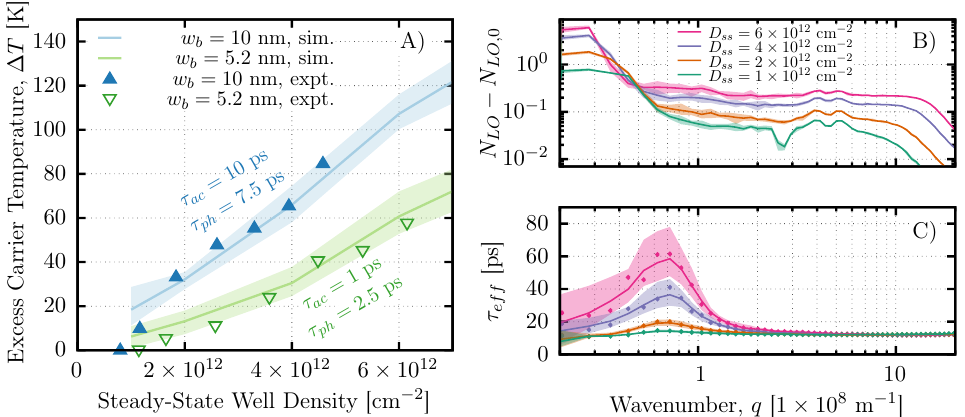}
	\caption{A) shows the best fits for the 10 nm and 5.2 nm barrier width. Subfigures B) and  C) show the $N_{LO}(q)$ and $\tau_{eff}(q)$ for the 10 nm barrier case for different steady-state well densities.}
\label{fig:wells}
\end{figure*}

With the inclusion of non-equilibrium acoustic phonons $\tau_{eff}$ becomes dependent on the steady-state well density (see Fig. \ref{fig:wells} C) since the peak around $0.7\times 10^8$ m$^{-1}$ is a result of the carrier-phonon interactions of the thermalized carriers in the well.  
As a result of this, $\tau_{eff}$ is also dependent on the laser power.
The model can then represent steeper $\Delta T$ vs. steady-state density, $D_{ss}$, relationships than the old relaxation-time methods.

\section{Discussion}
\label{sec:discussion}
The above model provides a fundamentally different way of understanding the phonon bottleneck phenomenon in confined structures, the elevated carrier temperatures measured from both time-resolved and CW experiments, and the implications for the realization of hot carrier solar cells.
Using a detailed balance model for the energy conversion efficiency of a hot carrier solar cell,
Tsai found that optical phonon lifetimes in excess of 100 ps are required in order to appreciably increase the conversion efficiency beyond fundamental Shockley-Queisser limit \cite{tsai_effects_2018}. 
Similarly, a large $\tau_{LO}$ was required in prior work using both rate equation models \cite{hirst_hot_2014} as well as ensemble Monte Carlo simulation \cite{baranowski_monte_2025,hathwar_nonequilibrium_2019}
Such large values of $\tau_{LO}$ are well in excess of the experimentally measured values in III-V materials from time-resolved Raman scattering experiments \cite{ferry_non-equilibrium_2021,grann_nonequilibrium_1996}, with no observed $\tau_{LO}$ in III-V materials exceeding 10 ps. 

The explanation provided by the present model which includes both nonequilibrium LO and LA phonons may be observed from the characteristic plots of $\tau_{eff}$ shown \ref{fig:wells}. 
The peak effective lifetime can be seen to be well in excess of 60 ps in the range of phonon wavevectors which most strongly coupled to the thermalized carrier distribution responsible for the steady state temperature (in CW experiments).  
In contrast, the effective lifetime for phonons at short wavelengths which interact with the photoexcited carriers at high energy is much closer to the bare LO phonon lifetime measured in Raman experiments. This suggests an equally important role played by the acoustic modes in hot carrier cooling and potential realization of energy conversion devices based on such phenomena.

Previously, ``phononic gap'' materials have been investigated such as InN where the gap between the lowest optical frequency and the highest acoustic mode supresses the Klemens mechanism, potentially increasing the LO phonon lifetime \cite{aliberti_investigation_2010}.
The proposed coupled LO-LA bottleneck mechanism in the present work suggests modification of the acoustic modes and their relaxation is equally important.  Conibeer et al.  proposed the introduction of gaps in the acoustic mode spectrum at half the LO phonon energy in order to suppress the Klemens process using a phononic crystal approach through a periodic structure such as semiconductor superlattice, or more effectively, a periodic array of quantum dots\cite{conibeer_slowing_2008}.  This possibility was more recently explored by Garg et al. in III-V superlattice structures\cite{garg_phonon_2020}, where the effect of the superlattice structure on the optical phonon anharmonic scattering rates was investigated. The present work suggests reduction of the acoustic mode decay time, through e.g. an artificial superlattice structure, could be equally effective in reducing the cooling rate for electrons and holes.

\section{Conclusion}
\label{sec:conclusion}

In this work we proposed an improved theoretical model over previous work\cite{baranowski_monte_2025} related to nonequilibrium optical phonon effects on hot carrier relaxation, that reveals a more complicated relationship between laser intensity and carrier thermalization through consideration of nonequilibrium acoustic phonons in addition to LO phonons.
The additional LA phonon blockade presents an alternative path for a hot carrier solar cell absorber layer with more relaxed design requirements than the conventionally proposed ones by phonon band engineering.

Future work will seek to quantify the effect of 3-phonon processes in the acoustic branches, wherein the LA phonon branches can interact with the TA phonon branches. 
This would allow for the description of the broadening of the daughter acoustic states.

\bibliography{ref.bib}

\begin{thebibliography}{37}%
\makeatletter
\providecommand \@ifxundefined [1]{%
 \@ifx{#1\undefined}
}%
\providecommand \@ifnum [1]{%
 \ifnum #1\expandafter \@firstoftwo
 \else \expandafter \@secondoftwo
 \fi
}%
\providecommand \@ifx [1]{%
 \ifx #1\expandafter \@firstoftwo
 \else \expandafter \@secondoftwo
 \fi
}%
\providecommand \natexlab [1]{#1}%
\providecommand \enquote  [1]{``#1''}%
\providecommand \bibnamefont  [1]{#1}%
\providecommand \bibfnamefont [1]{#1}%
\providecommand \citenamefont [1]{#1}%
\providecommand \href@noop [0]{\@secondoftwo}%
\providecommand \href [0]{\begingroup \@sanitize@url \@href}%
\providecommand \@href[1]{\@@startlink{#1}\@@href}%
\providecommand \@@href[1]{\endgroup#1\@@endlink}%
\providecommand \@sanitize@url [0]{\catcode `\\12\catcode `\$12\catcode
  `\&12\catcode `\#12\catcode `\^12\catcode `\_12\catcode `\%12\relax}%
\providecommand \@@startlink[1]{}%
\providecommand \@@endlink[0]{}%
\providecommand \url  [0]{\begingroup\@sanitize@url \@url }%
\providecommand \@url [1]{\endgroup\@href {#1}{\urlprefix }}%
\providecommand \urlprefix  [0]{URL }%
\providecommand \Eprint [0]{\href }%
\providecommand \doibase [0]{http://dx.doi.org/}%
\providecommand \selectlanguage [0]{\@gobble}%
\providecommand \bibinfo  [0]{\@secondoftwo}%
\providecommand \bibfield  [0]{\@secondoftwo}%
\providecommand \translation [1]{[#1]}%
\providecommand \BibitemOpen [0]{}%
\providecommand \bibitemStop [0]{}%
\providecommand \bibitemNoStop [0]{.\EOS\space}%
\providecommand \EOS [0]{\spacefactor3000\relax}%
\providecommand \BibitemShut  [1]{\csname bibitem#1\endcsname}%
\let\auto@bib@innerbib\@empty
%</preamble>
\bibitem [{\citenamefont {Zhang}\ \emph {et~al.}(2022)\citenamefont {Zhang},
  \citenamefont {Conibeer}, \citenamefont {Liu}, \citenamefont {Zhang},\ and\
  \citenamefont {Guillemoles}}]{zhang_review_2022}%
  \BibitemOpen
  \bibfield  {author} {\bibinfo {author} {\bibfnamefont {Y.}~\bibnamefont
  {Zhang}}, \bibinfo {author} {\bibfnamefont {G.}~\bibnamefont {Conibeer}},
  \bibinfo {author} {\bibfnamefont {S.}~\bibnamefont {Liu}}, \bibinfo {author}
  {\bibfnamefont {J.}~\bibnamefont {Zhang}}, \ and\ \bibinfo {author}
  {\bibfnamefont {J.-F.}\ \bibnamefont {Guillemoles}},\ }\bibfield  {title}
  {{\selectlanguage {english}\enquote {\bibinfo {title} {Review of the
  mechanisms for the phonon bottleneck effect in {III}–{V} semiconductors and
  their application for efficient hot carrier solar cells},}\ }}\href {\doibase
  10.1002/pip.3557} {\bibfield  {journal} {\bibinfo  {journal} {Progress in
  Photovoltaics: Research and Applications}\ }\textbf {\bibinfo {volume}
  {30}},\ \bibinfo {pages} {581--596} (\bibinfo {year} {2022})},\ \bibinfo
  {note} {\_eprint:
  https://onlinelibrary.wiley.com/doi/pdf/10.1002/pip.3557}\BibitemShut
  {NoStop}%
\bibitem [{\citenamefont {Kahmann}\ and\ \citenamefont
  {A. Loi}(2019)}]{kahmann_hot_2019}%
  \BibitemOpen
  \bibfield  {author} {\bibinfo {author} {\bibfnamefont {S.}~\bibnamefont
  {Kahmann}}\ and\ \bibinfo {author} {\bibfnamefont {M.}~\bibnamefont
  {A. Loi}},\ }\bibfield  {title} {{\selectlanguage {english}\enquote
  {\bibinfo {title} {Hot carrier solar cells and the potential of perovskites
  for breaking the {Shockley}–{Queisser} limit},}\ }}\href {\doibase
  10.1039/C8TC04641G} {\bibfield  {journal} {\bibinfo  {journal} {Journal of
  Materials Chemistry C}\ }\textbf {\bibinfo {volume} {7}},\ \bibinfo {pages}
  {2471--2486} (\bibinfo {year} {2019})},\ \bibinfo {note} {publisher: Royal
  Society of Chemistry}\BibitemShut {NoStop}%
\bibitem [{\citenamefont {Yang}\ \emph {et~al.}(2017)\citenamefont {Yang},
  \citenamefont {Wen}, \citenamefont {Xia}, \citenamefont {Sheng},
  \citenamefont {Ma}, \citenamefont {Kim}, \citenamefont {Tapping},
  \citenamefont {Harada}, \citenamefont {Kee}, \citenamefont {Huang},
  \citenamefont {Cheng}, \citenamefont {Green}, \citenamefont {Ho-Baillie},
  \citenamefont {Huang}, \citenamefont {Shrestha}, \citenamefont {Patterson},\
  and\ \citenamefont {Conibeer}}]{yang_acoustic-optical_2017}%
  \BibitemOpen
  \bibfield  {author} {\bibinfo {author} {\bibfnamefont {J.}~\bibnamefont
  {Yang}}, \bibinfo {author} {\bibfnamefont {X.}~\bibnamefont {Wen}}, \bibinfo
  {author} {\bibfnamefont {H.}~\bibnamefont {Xia}}, \bibinfo {author}
  {\bibfnamefont {R.}~\bibnamefont {Sheng}}, \bibinfo {author} {\bibfnamefont
  {Q.}~\bibnamefont {Ma}}, \bibinfo {author} {\bibfnamefont {J.}~\bibnamefont
  {Kim}}, \bibinfo {author} {\bibfnamefont {P.}~\bibnamefont {Tapping}},
  \bibinfo {author} {\bibfnamefont {T.}~\bibnamefont {Harada}}, \bibinfo
  {author} {\bibfnamefont {T.~W.}\ \bibnamefont {Kee}}, \bibinfo {author}
  {\bibfnamefont {F.}~\bibnamefont {Huang}}, \bibinfo {author} {\bibfnamefont
  {Y.-B.}\ \bibnamefont {Cheng}}, \bibinfo {author} {\bibfnamefont
  {M.}~\bibnamefont {Green}}, \bibinfo {author} {\bibfnamefont
  {A.}~\bibnamefont {Ho-Baillie}}, \bibinfo {author} {\bibfnamefont
  {S.}~\bibnamefont {Huang}}, \bibinfo {author} {\bibfnamefont
  {S.}~\bibnamefont {Shrestha}}, \bibinfo {author} {\bibfnamefont
  {R.}~\bibnamefont {Patterson}}, \ and\ \bibinfo {author} {\bibfnamefont
  {G.}~\bibnamefont {Conibeer}},\ }\bibfield  {title} {{\selectlanguage
  {english}\enquote {\bibinfo {title} {Acoustic-optical phonon up-conversion
  and hot-phonon bottleneck in lead-halide perovskites},}\ }}\href {\doibase
  10.1038/ncomms14120} {\bibfield  {journal} {\bibinfo  {journal} {Nature
  Communications}\ }\textbf {\bibinfo {volume} {8}},\ \bibinfo {pages} {14120}
  (\bibinfo {year} {2017})},\ \bibinfo {note} {number: 1 Publisher: Nature
  Publishing Group}\BibitemShut {NoStop}%
\bibitem [{\citenamefont {Ferry}\ \emph {et~al.}(2020)\citenamefont {Ferry},
  \citenamefont {Goodnick}, \citenamefont {Whiteside},\ and\ \citenamefont
  {Sellers}}]{ferry_challenges_2020}%
  \BibitemOpen
  \bibfield  {author} {\bibinfo {author} {\bibfnamefont {D.~K.}\ \bibnamefont
  {Ferry}}, \bibinfo {author} {\bibfnamefont {S.~M.}\ \bibnamefont {Goodnick}},
  \bibinfo {author} {\bibfnamefont {V.~R.}\ \bibnamefont {Whiteside}}, \ and\
  \bibinfo {author} {\bibfnamefont {I.~R.}\ \bibnamefont {Sellers}},\
  }\bibfield  {title} {\enquote {\bibinfo {title} {Challenges, myths, and
  opportunities in hot carrier solar cells},}\ }\href {\doibase
  10.1063/5.0028981} {\bibfield  {journal} {\bibinfo  {journal} {Journal of
  Applied Physics}\ }\textbf {\bibinfo {volume} {128}},\ \bibinfo {pages}
  {220903} (\bibinfo {year} {2020})},\ \bibinfo {note} {publisher: American
  Institute of Physics}\BibitemShut {NoStop}%
\bibitem [{\citenamefont {Shockley}\ and\ \citenamefont
  {Queisser}(1961)}]{shockley_detailed_1961}%
  \BibitemOpen
  \bibfield  {author} {\bibinfo {author} {\bibfnamefont {W.}~\bibnamefont
  {Shockley}}\ and\ \bibinfo {author} {\bibfnamefont {H.~J.}\ \bibnamefont
  {Queisser}},\ }\bibfield  {title} {\enquote {\bibinfo {title} {Detailed
  {Balance} {Limit} of {Efficiency} of p‐n {Junction} {Solar} {Cells}},}\
  }\href {\doibase 10.1063/1.1736034} {\bibfield  {journal} {\bibinfo
  {journal} {Journal of Applied Physics}\ }\textbf {\bibinfo {volume} {32}},\
  \bibinfo {pages} {510--519} (\bibinfo {year} {1961})},\ \bibinfo {note}
  {publisher: American Institute of Physics}\BibitemShut {NoStop}%
\bibitem [{\citenamefont {Ross}\ and\ \citenamefont
  {Nozik}(1982)}]{ross_efficiency_1982}%
  \BibitemOpen
  \bibfield  {author} {\bibinfo {author} {\bibfnamefont {R.~T.}\ \bibnamefont
  {Ross}}\ and\ \bibinfo {author} {\bibfnamefont {A.~J.}\ \bibnamefont
  {Nozik}},\ }\bibfield  {title} {\enquote {\bibinfo {title} {Efficiency of
  hot‐carrier solar energy converters},}\ }\href {\doibase 10.1063/1.331124}
  {\bibfield  {journal} {\bibinfo  {journal} {Journal of Applied Physics}\
  }\textbf {\bibinfo {volume} {53}},\ \bibinfo {pages} {3813--3818} (\bibinfo
  {year} {1982})},\ \bibinfo {note} {publisher: American Institute of
  Physics}\BibitemShut {NoStop}%
\bibitem [{\citenamefont {Zou}\ \emph {et~al.}(2023)\citenamefont {Zou},
  \citenamefont {Esmaielpour}, \citenamefont {Suchet}, \citenamefont
  {Guillemoles},\ and\ \citenamefont {Goodnick}}]{zou_role_2023}%
  \BibitemOpen
  \bibfield  {author} {\bibinfo {author} {\bibfnamefont {Y.}~\bibnamefont
  {Zou}}, \bibinfo {author} {\bibfnamefont {H.}~\bibnamefont {Esmaielpour}},
  \bibinfo {author} {\bibfnamefont {D.}~\bibnamefont {Suchet}}, \bibinfo
  {author} {\bibfnamefont {J.-F.}\ \bibnamefont {Guillemoles}}, \ and\ \bibinfo
  {author} {\bibfnamefont {S.~M.}\ \bibnamefont {Goodnick}},\ }\bibfield
  {title} {{\selectlanguage {english}\enquote {\bibinfo {title} {The role of
  nonequilibrium {LO} phonons, {Pauli} exclusion, and intervalley pathways on
  the relaxation of hot carriers in {InGaAs}/{InGaAsP} multi-quantum-wells},}\
  }}\href {\doibase 10.1038/s41598-023-32125-2} {\bibfield  {journal} {\bibinfo
   {journal} {Scientific Reports}\ }\textbf {\bibinfo {volume} {13}},\ \bibinfo
  {pages} {5601} (\bibinfo {year} {2023})},\ \bibinfo {note} {number: 1
  Publisher: Nature Publishing Group}\BibitemShut {NoStop}%
\bibitem [{\citenamefont {Baranowski}\ \emph {et~al.}(2025)\citenamefont
  {Baranowski}, \citenamefont {Sellers}, \citenamefont {Vasileska},\ and\
  \citenamefont {Goodnick}}]{baranowski_monte_2025}%
  \BibitemOpen
  \bibfield  {author} {\bibinfo {author} {\bibfnamefont {I.}~\bibnamefont
  {Baranowski}}, \bibinfo {author} {\bibfnamefont {I.~R.}\ \bibnamefont
  {Sellers}}, \bibinfo {author} {\bibfnamefont {D.}~\bibnamefont {Vasileska}},
  \ and\ \bibinfo {author} {\bibfnamefont {S.~M.}\ \bibnamefont {Goodnick}},\
  }\bibfield  {title} {\enquote {\bibinfo {title} {Monte {Carlo} simulation of
  ultrafast carrier relaxation in type-{II} {MQW} system},}\ }\href {\doibase
  10.1117/1.JPE.15.012505} {\bibfield  {journal} {\bibinfo  {journal} {Journal
  of Photonics for Energy}\ }\textbf {\bibinfo {volume} {15}},\ \bibinfo
  {pages} {012505} (\bibinfo {year} {2025})},\ \bibinfo {note} {publisher:
  SPIE}\BibitemShut {NoStop}%
\bibitem [{\citenamefont {Faber}, \citenamefont {Filipovic},\ and\
  \citenamefont {Koster}(2024)}]{faber_hot_2024}%
  \BibitemOpen
  \bibfield  {author} {\bibinfo {author} {\bibfnamefont {T.}~\bibnamefont
  {Faber}}, \bibinfo {author} {\bibfnamefont {L.}~\bibnamefont {Filipovic}}, \
  and\ \bibinfo {author} {\bibfnamefont {L.}~\bibnamefont {Koster}},\
  }\bibfield  {title} {\enquote {\bibinfo {title} {The {Hot} {Phonon}
  {Bottleneck} {Effect} in {Metal} {Halide} {Perovskites}},}\ }\href {\doibase
  10.1021/acs.jpclett.4c03133} {\bibfield  {journal} {\bibinfo  {journal} {The
  Journal of Physical Chemistry Letters}\ }\textbf {\bibinfo {volume} {15}},\
  \bibinfo {pages} {12601--12607} (\bibinfo {year} {2024})},\ \bibinfo {note}
  {publisher: American Chemical Society}\BibitemShut {NoStop}%
\bibitem [{\citenamefont {Esmaielpour}\ \emph {et~al.}(2021)\citenamefont
  {Esmaielpour}, \citenamefont {Durant}, \citenamefont {Dorman}, \citenamefont
  {Whiteside}, \citenamefont {Garg}, \citenamefont {Mishima}, \citenamefont
  {Santos}, \citenamefont {Sellers}, \citenamefont {Guillemoles},\ and\
  \citenamefont {Suchet}}]{esmaielpour_hot_2021}%
  \BibitemOpen
  \bibfield  {author} {\bibinfo {author} {\bibfnamefont {H.}~\bibnamefont
  {Esmaielpour}}, \bibinfo {author} {\bibfnamefont {B.~K.}\ \bibnamefont
  {Durant}}, \bibinfo {author} {\bibfnamefont {K.~R.}\ \bibnamefont {Dorman}},
  \bibinfo {author} {\bibfnamefont {V.~R.}\ \bibnamefont {Whiteside}}, \bibinfo
  {author} {\bibfnamefont {J.}~\bibnamefont {Garg}}, \bibinfo {author}
  {\bibfnamefont {T.~D.}\ \bibnamefont {Mishima}}, \bibinfo {author}
  {\bibfnamefont {M.~B.}\ \bibnamefont {Santos}}, \bibinfo {author}
  {\bibfnamefont {I.~R.}\ \bibnamefont {Sellers}}, \bibinfo {author}
  {\bibfnamefont {J.-F.}\ \bibnamefont {Guillemoles}}, \ and\ \bibinfo {author}
  {\bibfnamefont {D.}~\bibnamefont {Suchet}},\ }\bibfield  {title} {\enquote
  {\bibinfo {title} {Hot carrier relaxation and inhibited thermalization in
  superlattice heterostructures: {The} potential for phonon management},}\
  }\href {\doibase 10.1063/5.0052600} {\bibfield  {journal} {\bibinfo
  {journal} {Applied Physics Letters}\ }\textbf {\bibinfo {volume} {118}},\
  \bibinfo {pages} {213902} (\bibinfo {year} {2021})},\ \bibinfo {note}
  {publisher: American Institute of Physics}\BibitemShut {NoStop}%
\bibitem [{\citenamefont {Hejda}\ and\ \citenamefont
  {Král}(1993)}]{hejda_hot-electron_1993}%
  \BibitemOpen
  \bibfield  {author} {\bibinfo {author} {\bibfnamefont {B.}~\bibnamefont
  {Hejda}}\ and\ \bibinfo {author} {\bibfnamefont {K.}~\bibnamefont {Král}},\
  }\bibfield  {title} {\enquote {\bibinfo {title} {Hot-electron cooling and
  second-generation phonons in polar semiconductors},}\ }\href {\doibase
  10.1103/PhysRevB.47.15554} {\bibfield  {journal} {\bibinfo  {journal}
  {Physical Review B}\ }\textbf {\bibinfo {volume} {47}},\ \bibinfo {pages}
  {15554--15561} (\bibinfo {year} {1993})},\ \bibinfo {note} {publisher:
  American Physical Society}\BibitemShut {NoStop}%
\bibitem [{\citenamefont {Juršėnas}, \citenamefont {Kurilčik},\ and\
  \citenamefont {Žukauskas}(1998)}]{jursenas_dense_1998}%
  \BibitemOpen
  \bibfield  {author} {\bibinfo {author} {\bibfnamefont {S.}~\bibnamefont
  {Juršėnas}}, \bibinfo {author} {\bibfnamefont {G.}~\bibnamefont
  {Kurilčik}}, \ and\ \bibinfo {author} {\bibfnamefont {A.}~\bibnamefont
  {Žukauskas}},\ }\bibfield  {title} {\enquote {\bibinfo {title} {Dense
  electron-hole plasma cooling due to second nonequilibrium-phonon bottleneck
  in {CdS} crystallites},}\ }\href {\doibase 10.1103/PhysRevB.58.12937}
  {\bibfield  {journal} {\bibinfo  {journal} {Physical Review B}\ }\textbf
  {\bibinfo {volume} {58}},\ \bibinfo {pages} {12937--12943} (\bibinfo {year}
  {1998})},\ \bibinfo {note} {publisher: American Physical Society}\BibitemShut
  {NoStop}%
\bibitem [{\citenamefont {Juršenas}, \citenamefont {Kurilčik},\ and\
  \citenamefont {Žukauskas}(1999)}]{jursenas_carrier_1999}%
  \BibitemOpen
  \bibfield  {author} {\bibinfo {author} {\bibfnamefont {S.}~\bibnamefont
  {Juršenas}}, \bibinfo {author} {\bibfnamefont {G.}~\bibnamefont
  {Kurilčik}}, \ and\ \bibinfo {author} {\bibfnamefont {A.}~\bibnamefont
  {Žukauskas}},\ }\bibfield  {title} {{\selectlanguage {english}\enquote
  {\bibinfo {title} {Carrier cooling in {CdS} crystallites under extremely high
  photoexcitation},}\ }}\href {\doibase 10.1238/Physica.Topical.079a00167}
  {\bibfield  {journal} {\bibinfo  {journal} {Physica Scripta}\ }\textbf
  {\bibinfo {volume} {1999}},\ \bibinfo {pages} {167} (\bibinfo {year}
  {1999})},\ \bibinfo {note} {publisher: IOP Publishing}\BibitemShut {NoStop}%
\bibitem [{\citenamefont
  {Král}(1994{\natexlab{a}})}]{kral_electrophonon_1994}%
  \BibitemOpen
  \bibfield  {author} {\bibinfo {author} {\bibfnamefont {K.}~\bibnamefont
  {Král}},\ }\bibfield  {title} {\enquote {\bibinfo {title} {Electrophonon
  resonance in
  {${\mathrm{{Al}}}_{\mathit{x}}$}{${\mathrm{{Ga}}}_{1\mathrm{\ensuremath{-}}\mathit{x}}$}{As}-{GaAs}
  quasi-two-dimensional quantum wells},}\ }\href {\doibase
  10.1103/PhysRevB.50.7640} {\bibfield  {journal} {\bibinfo  {journal}
  {Physical Review B}\ }\textbf {\bibinfo {volume} {50}},\ \bibinfo {pages}
  {7640--7654} (\bibinfo {year} {1994}{\natexlab{a}})},\ \bibinfo {note}
  {publisher: American Physical Society}\BibitemShut {NoStop}%
\bibitem [{\citenamefont {Král}(1994{\natexlab{b}})}]{kral_hot-electron_1994}%
  \BibitemOpen
  \bibfield  {author} {\bibinfo {author} {\bibfnamefont {K.}~\bibnamefont
  {Král}},\ }\bibfield  {title} {\enquote {\bibinfo {title} {Hot-electron
  cooling and hot-phonon generation with collision broadening},}\ }\href
  {\doibase 10.1103/PhysRevB.50.7988} {\bibfield  {journal} {\bibinfo
  {journal} {Physical Review B}\ }\textbf {\bibinfo {volume} {50}},\ \bibinfo
  {pages} {7988--7991} (\bibinfo {year} {1994}{\natexlab{b}})},\ \bibinfo
  {note} {publisher: American Physical Society}\BibitemShut {NoStop}%
\bibitem [{\citenamefont {Usher}\ and\ \citenamefont
  {Srivastava}(1994)}]{usher_theoretical_1994}%
  \BibitemOpen
  \bibfield  {author} {\bibinfo {author} {\bibfnamefont {S.}~\bibnamefont
  {Usher}}\ and\ \bibinfo {author} {\bibfnamefont {G.~P.}\ \bibnamefont
  {Srivastava}},\ }\bibfield  {title} {\enquote {\bibinfo {title} {Theoretical
  study of the anharmonic decay of nonequilibrium {LO} phonons in semiconductor
  structures},}\ }\href {\doibase 10.1103/PhysRevB.50.14179} {\bibfield
  {journal} {\bibinfo  {journal} {Physical Review B}\ }\textbf {\bibinfo
  {volume} {50}},\ \bibinfo {pages} {14179--14186} (\bibinfo {year} {1994})},\
  \bibinfo {note} {publisher: American Physical Society}\BibitemShut {NoStop}%
\bibitem [{\citenamefont {Žukauskas}(1998)}]{zukauskas_second_1998}%
  \BibitemOpen
  \bibfield  {author} {\bibinfo {author} {\bibfnamefont {A.}~\bibnamefont
  {Žukauskas}},\ }\bibfield  {title} {\enquote {\bibinfo {title} {Second
  nonequilibrium-phonon bottleneck for carrier cooling in highly excited polar
  semiconductors},}\ }\href {\doibase 10.1103/PhysRevB.57.15337} {\bibfield
  {journal} {\bibinfo  {journal} {Physical Review B}\ }\textbf {\bibinfo
  {volume} {57}},\ \bibinfo {pages} {15337--15344} (\bibinfo {year} {1998})},\
  \bibinfo {note} {publisher: American Physical Society}\BibitemShut {NoStop}%
\bibitem [{\citenamefont {Dyson}\ and\ \citenamefont
  {Ridley}(2013)}]{dyson_role_2013}%
  \BibitemOpen
  \bibfield  {author} {\bibinfo {author} {\bibfnamefont {A.}~\bibnamefont
  {Dyson}}\ and\ \bibinfo {author} {\bibfnamefont {B.~K.}\ \bibnamefont
  {Ridley}},\ }\bibfield  {title} {\enquote {\bibinfo {title} {The role of the
  products of the decay of optical phonons},}\ }\href {\doibase
  10.1063/1.4790280} {\bibfield  {journal} {\bibinfo  {journal} {Applied
  Physics Letters}\ }\textbf {\bibinfo {volume} {102}},\ \bibinfo {pages}
  {042108} (\bibinfo {year} {2013})}\BibitemShut {NoStop}%
\bibitem [{\citenamefont {Grann}, \citenamefont {Tsen},\ and\ \citenamefont
  {Ferry}(1996)}]{grann_nonequilibrium_1996}%
  \BibitemOpen
  \bibfield  {author} {\bibinfo {author} {\bibfnamefont {E.~D.}\ \bibnamefont
  {Grann}}, \bibinfo {author} {\bibfnamefont {K.~T.}\ \bibnamefont {Tsen}}, \
  and\ \bibinfo {author} {\bibfnamefont {D.~K.}\ \bibnamefont {Ferry}},\
  }\bibfield  {title} {\enquote {\bibinfo {title} {Nonequilibrium phonon
  dynamics and electron distribution functions in {InP} and {InAs}},}\ }\href
  {\doibase 10.1103/PhysRevB.53.9847} {\bibfield  {journal} {\bibinfo
  {journal} {Physical Review B}\ }\textbf {\bibinfo {volume} {53}},\ \bibinfo
  {pages} {9847--9851} (\bibinfo {year} {1996})},\ \bibinfo {note} {publisher:
  American Physical Society}\BibitemShut {NoStop}%
\bibitem [{\citenamefont {Hirst}\ \emph {et~al.}(2011)\citenamefont {Hirst},
  \citenamefont {Führer}, \citenamefont {Farrell}, \citenamefont {LeBris},
  \citenamefont {Guillemoles}, \citenamefont {Tayebjee}, \citenamefont {Clady},
  \citenamefont {Schmidt}, \citenamefont {Wang}, \citenamefont {Sugiyama},\
  and\ \citenamefont {Ekins-Daukes}}]{hirst_hot_2011}%
  \BibitemOpen
  \bibfield  {author} {\bibinfo {author} {\bibfnamefont {L.}~\bibnamefont
  {Hirst}}, \bibinfo {author} {\bibfnamefont {M.}~\bibnamefont {Führer}},
  \bibinfo {author} {\bibfnamefont {D.~J.}\ \bibnamefont {Farrell}}, \bibinfo
  {author} {\bibfnamefont {A.}~\bibnamefont {LeBris}}, \bibinfo {author}
  {\bibfnamefont {J.-F.}\ \bibnamefont {Guillemoles}}, \bibinfo {author}
  {\bibfnamefont {M.~J.~Y.}\ \bibnamefont {Tayebjee}}, \bibinfo {author}
  {\bibfnamefont {R.}~\bibnamefont {Clady}}, \bibinfo {author} {\bibfnamefont
  {T.~W.}\ \bibnamefont {Schmidt}}, \bibinfo {author} {\bibfnamefont
  {Y.}~\bibnamefont {Wang}}, \bibinfo {author} {\bibfnamefont {M.}~\bibnamefont
  {Sugiyama}}, \ and\ \bibinfo {author} {\bibfnamefont {N.~J.}\ \bibnamefont
  {Ekins-Daukes}},\ }\bibfield  {title} {\enquote {\bibinfo {title} {Hot
  carrier dynamics in {InGaAs}/{GaAsP} quantum well solar cells},}\ }in\ \href
  {\doibase 10.1109/PVSC.2011.6186643} {\emph {\bibinfo {booktitle} {2011 37th
  {IEEE} {Photovoltaic} {Specialists} {Conference}}}}\ (\bibinfo {year}
  {2011})\ pp.\ \bibinfo {pages} {003302--003306},\ \bibinfo {note} {iSSN:
  0160-8371}\BibitemShut {NoStop}%
\bibitem [{\citenamefont {Esmaielpour}\ \emph {et~al.}(2018)\citenamefont
  {Esmaielpour}, \citenamefont {Whiteside}, \citenamefont {Piyathilaka},
  \citenamefont {Vijeyaragunathan}, \citenamefont {Wang}, \citenamefont
  {Adcock-Smith}, \citenamefont {Roberts}, \citenamefont {Mishima},
  \citenamefont {Santos}, \citenamefont {Bristow},\ and\ \citenamefont
  {Sellers}}]{esmaielpour_enhanced_2018}%
  \BibitemOpen
  \bibfield  {author} {\bibinfo {author} {\bibfnamefont {H.}~\bibnamefont
  {Esmaielpour}}, \bibinfo {author} {\bibfnamefont {V.~R.}\ \bibnamefont
  {Whiteside}}, \bibinfo {author} {\bibfnamefont {H.~P.}\ \bibnamefont
  {Piyathilaka}}, \bibinfo {author} {\bibfnamefont {S.}~\bibnamefont
  {Vijeyaragunathan}}, \bibinfo {author} {\bibfnamefont {B.}~\bibnamefont
  {Wang}}, \bibinfo {author} {\bibfnamefont {E.}~\bibnamefont {Adcock-Smith}},
  \bibinfo {author} {\bibfnamefont {K.~P.}\ \bibnamefont {Roberts}}, \bibinfo
  {author} {\bibfnamefont {T.~D.}\ \bibnamefont {Mishima}}, \bibinfo {author}
  {\bibfnamefont {M.~B.}\ \bibnamefont {Santos}}, \bibinfo {author}
  {\bibfnamefont {A.~D.}\ \bibnamefont {Bristow}}, \ and\ \bibinfo {author}
  {\bibfnamefont {I.~R.}\ \bibnamefont {Sellers}},\ }\bibfield  {title}
  {{\selectlanguage {english}\enquote {\bibinfo {title} {Enhanced hot electron
  lifetimes in quantum wells with inhibited phonon coupling},}\ }}\href
  {\doibase 10.1038/s41598-018-30894-9} {\bibfield  {journal} {\bibinfo
  {journal} {Scientific Reports}\ }\textbf {\bibinfo {volume} {8}},\ \bibinfo
  {pages} {12473} (\bibinfo {year} {2018})},\ \bibinfo {note} {number: 1
  Publisher: Nature Publishing Group}\BibitemShut {NoStop}%
\bibitem [{\citenamefont {Sandner}\ \emph {et~al.}(2023)\citenamefont
  {Sandner}, \citenamefont {Esmaielpour}, \citenamefont {Giudice},
  \citenamefont {Meder}, \citenamefont {Nuber}, \citenamefont {Kienberger},
  \citenamefont {Koblmüller},\ and\ \citenamefont {Iglev}}]{sandner_hot_2023}%
  \BibitemOpen
  \bibfield  {author} {\bibinfo {author} {\bibfnamefont {D.}~\bibnamefont
  {Sandner}}, \bibinfo {author} {\bibfnamefont {H.}~\bibnamefont
  {Esmaielpour}}, \bibinfo {author} {\bibfnamefont {F.~d.}\ \bibnamefont
  {Giudice}}, \bibinfo {author} {\bibfnamefont {S.}~\bibnamefont {Meder}},
  \bibinfo {author} {\bibfnamefont {M.}~\bibnamefont {Nuber}}, \bibinfo
  {author} {\bibfnamefont {R.}~\bibnamefont {Kienberger}}, \bibinfo {author}
  {\bibfnamefont {G.}~\bibnamefont {Koblmüller}}, \ and\ \bibinfo {author}
  {\bibfnamefont {H.}~\bibnamefont {Iglev}},\ }\bibfield  {title} {\enquote
  {\bibinfo {title} {Hot {Electron} {Dynamics} in {InAs}–{AlAsSb}
  {Core}–{Shell} {Nanowires}},}\ }\href {\doibase 10.1021/acsaem.3c01565}
  {\bibfield  {journal} {\bibinfo  {journal} {ACS Applied Energy Materials}\
  }\textbf {\bibinfo {volume} {6}},\ \bibinfo {pages} {10467--10474} (\bibinfo
  {year} {2023})},\ \bibinfo {note} {publisher: American Chemical
  Society}\BibitemShut {NoStop}%
\bibitem [{\citenamefont {Zhang}\ \emph {et~al.}(2016)\citenamefont {Zhang},
  \citenamefont {Tayebjee}, \citenamefont {Smyth}, \citenamefont {Dvořák},
  \citenamefont {Wen}, \citenamefont {Xia}, \citenamefont {Heilmann},
  \citenamefont {Liao}, \citenamefont {Zhang}, \citenamefont {Williamson},
  \citenamefont {Williams}, \citenamefont {Bremner}, \citenamefont {Shrestha},
  \citenamefont {Huang}, \citenamefont {Schmidt},\ and\ \citenamefont
  {Conibeer}}]{zhang_extended_2016}%
  \BibitemOpen
  \bibfield  {author} {\bibinfo {author} {\bibfnamefont {Y.}~\bibnamefont
  {Zhang}}, \bibinfo {author} {\bibfnamefont {M.~J.~Y.}\ \bibnamefont
  {Tayebjee}}, \bibinfo {author} {\bibfnamefont {S.}~\bibnamefont {Smyth}},
  \bibinfo {author} {\bibfnamefont {M.}~\bibnamefont {Dvořák}}, \bibinfo
  {author} {\bibfnamefont {X.}~\bibnamefont {Wen}}, \bibinfo {author}
  {\bibfnamefont {H.}~\bibnamefont {Xia}}, \bibinfo {author} {\bibfnamefont
  {M.}~\bibnamefont {Heilmann}}, \bibinfo {author} {\bibfnamefont
  {Y.}~\bibnamefont {Liao}}, \bibinfo {author} {\bibfnamefont {Z.}~\bibnamefont
  {Zhang}}, \bibinfo {author} {\bibfnamefont {T.}~\bibnamefont {Williamson}},
  \bibinfo {author} {\bibfnamefont {J.}~\bibnamefont {Williams}}, \bibinfo
  {author} {\bibfnamefont {S.}~\bibnamefont {Bremner}}, \bibinfo {author}
  {\bibfnamefont {S.}~\bibnamefont {Shrestha}}, \bibinfo {author}
  {\bibfnamefont {S.}~\bibnamefont {Huang}}, \bibinfo {author} {\bibfnamefont
  {T.~W.}\ \bibnamefont {Schmidt}}, \ and\ \bibinfo {author} {\bibfnamefont
  {G.~J.}\ \bibnamefont {Conibeer}},\ }\bibfield  {title} {\enquote {\bibinfo
  {title} {Extended hot carrier lifetimes observed in bulk
  {In0}.265±0.{02Ga0}.{735N} under high-density photoexcitation},}\ }\href
  {\doibase 10.1063/1.4945594} {\bibfield  {journal} {\bibinfo  {journal}
  {Applied Physics Letters}\ }\textbf {\bibinfo {volume} {108}},\ \bibinfo
  {pages} {131904} (\bibinfo {year} {2016})},\ \bibinfo {note} {publisher:
  American Institute of Physics}\BibitemShut {NoStop}%
\bibitem [{\citenamefont {Piyathilaka}\ \emph {et~al.}(2021)\citenamefont
  {Piyathilaka}, \citenamefont {Sooriyagoda}, \citenamefont {Esmaielpour},
  \citenamefont {Whiteside}, \citenamefont {Mishima}, \citenamefont {Santos},
  \citenamefont {Sellers},\ and\ \citenamefont
  {Bristow}}]{piyathilaka_hot-carrier_2021}%
  \BibitemOpen
  \bibfield  {author} {\bibinfo {author} {\bibfnamefont {H.~P.}\ \bibnamefont
  {Piyathilaka}}, \bibinfo {author} {\bibfnamefont {R.}~\bibnamefont
  {Sooriyagoda}}, \bibinfo {author} {\bibfnamefont {H.}~\bibnamefont
  {Esmaielpour}}, \bibinfo {author} {\bibfnamefont {V.~R.}\ \bibnamefont
  {Whiteside}}, \bibinfo {author} {\bibfnamefont {T.~D.}\ \bibnamefont
  {Mishima}}, \bibinfo {author} {\bibfnamefont {M.~B.}\ \bibnamefont {Santos}},
  \bibinfo {author} {\bibfnamefont {I.~R.}\ \bibnamefont {Sellers}}, \ and\
  \bibinfo {author} {\bibfnamefont {A.~D.}\ \bibnamefont {Bristow}},\
  }\bibfield  {title} {{\selectlanguage {english}\enquote {\bibinfo {title}
  {Hot-carrier dynamics in {InAs}/{AlAsSb} multiple-quantum wells},}\ }}\href
  {\doibase 10.1038/s41598-021-89815-y} {\bibfield  {journal} {\bibinfo
  {journal} {Scientific Reports}\ }\textbf {\bibinfo {volume} {11}},\ \bibinfo
  {pages} {10483} (\bibinfo {year} {2021})},\ \bibinfo {note} {number: 1
  Publisher: Nature Publishing Group}\BibitemShut {NoStop}%
\bibitem [{\citenamefont {Hathwar}\ \emph {et~al.}(2019)\citenamefont
  {Hathwar}, \citenamefont {Zou}, \citenamefont {Jirauschek},\ and\
  \citenamefont {Goodnick}}]{hathwar_nonequilibrium_2019}%
  \BibitemOpen
  \bibfield  {author} {\bibinfo {author} {\bibfnamefont {R.}~\bibnamefont
  {Hathwar}}, \bibinfo {author} {\bibfnamefont {Y.}~\bibnamefont {Zou}},
  \bibinfo {author} {\bibfnamefont {C.}~\bibnamefont {Jirauschek}}, \ and\
  \bibinfo {author} {\bibfnamefont {S.~M.}\ \bibnamefont {Goodnick}},\
  }\bibfield  {title} {{\selectlanguage {english}\enquote {\bibinfo {title}
  {Nonequilibrium electron and phonon dynamics in advanced concept solar
  cells},}\ }}\href {\doibase 10.1088/1361-6463/aaf750} {\bibfield  {journal}
  {\bibinfo  {journal} {Journal of Physics D: Applied Physics}\ }\textbf
  {\bibinfo {volume} {52}},\ \bibinfo {pages} {093001} (\bibinfo {year}
  {2019})},\ \bibinfo {note} {publisher: IOP Publishing}\BibitemShut {NoStop}%
\bibitem [{\citenamefont {Lugli}\ and\ \citenamefont
  {Goodnick}(1987)}]{lugli_nonequilibrium_1987}%
  \BibitemOpen
  \bibfield  {author} {\bibinfo {author} {\bibfnamefont {P.}~\bibnamefont
  {Lugli}}\ and\ \bibinfo {author} {\bibfnamefont {S.~M.}\ \bibnamefont
  {Goodnick}},\ }\bibfield  {title} {\enquote {\bibinfo {title} {Nonequilibrium
  longitudinal-optical phonon effects in {GaAs}-{AlGaAs} quantum wells},}\
  }\href {\doibase 10.1103/PhysRevLett.59.716} {\bibfield  {journal} {\bibinfo
  {journal} {Physical Review Letters}\ }\textbf {\bibinfo {volume} {59}},\
  \bibinfo {pages} {716--719} (\bibinfo {year} {1987})},\ \bibinfo {note}
  {publisher: American Physical Society}\BibitemShut {NoStop}%
\bibitem [{\citenamefont {Lugli}\ \emph {et~al.}(1989)\citenamefont {Lugli},
  \citenamefont {Bordone}, \citenamefont {Reggiani}, \citenamefont {Rieger},
  \citenamefont {Kocevar},\ and\ \citenamefont {Goodnick}}]{lugli_monte_1989}%
  \BibitemOpen
  \bibfield  {author} {\bibinfo {author} {\bibfnamefont {P.}~\bibnamefont
  {Lugli}}, \bibinfo {author} {\bibfnamefont {P.}~\bibnamefont {Bordone}},
  \bibinfo {author} {\bibfnamefont {L.}~\bibnamefont {Reggiani}}, \bibinfo
  {author} {\bibfnamefont {M.}~\bibnamefont {Rieger}}, \bibinfo {author}
  {\bibfnamefont {P.}~\bibnamefont {Kocevar}}, \ and\ \bibinfo {author}
  {\bibfnamefont {S.~M.}\ \bibnamefont {Goodnick}},\ }\bibfield  {title}
  {\enquote {\bibinfo {title} {Monte {Carlo} studies of nonequilibrium phonon
  effects in polar semiconductors and quantum wells. {I}. {Laser}
  photoexcitation},}\ }\href {\doibase 10.1103/PhysRevB.39.7852} {\bibfield
  {journal} {\bibinfo  {journal} {Physical Review B}\ }\textbf {\bibinfo
  {volume} {39}},\ \bibinfo {pages} {7852--7865} (\bibinfo {year} {1989})},\
  \bibinfo {note} {publisher: American Physical Society}\BibitemShut {NoStop}%
\bibitem [{\citenamefont {Baranowski}\ \emph {et~al.}(2023)\citenamefont
  {Baranowski}, \citenamefont {Zou}, \citenamefont {Esmaielpour}, \citenamefont
  {Sellers}, \citenamefont {Vasileska},\ and\ \citenamefont
  {Goodnick}}]{baranowski_monte_2023}%
  \BibitemOpen
  \bibfield  {author} {\bibinfo {author} {\bibfnamefont {I.}~\bibnamefont
  {Baranowski}}, \bibinfo {author} {\bibfnamefont {Y.}~\bibnamefont {Zou}},
  \bibinfo {author} {\bibfnamefont {H.}~\bibnamefont {Esmaielpour}}, \bibinfo
  {author} {\bibfnamefont {I.}~\bibnamefont {Sellers}}, \bibinfo {author}
  {\bibfnamefont {D.}~\bibnamefont {Vasileska}}, \ and\ \bibinfo {author}
  {\bibfnamefont {S.~M.}\ \bibnamefont {Goodnick}},\ }\bibfield  {title}
  {\enquote {\bibinfo {title} {Monte {Carlo} simulation of ultrafast carrier
  relaxation in type {I} and type {II} {InAs}-based quantum wells},}\ }in\
  \href {\doibase 10.1117/12.2656991} {\emph {\bibinfo {booktitle} {Physics,
  {Simulation}, and {Photonic} {Engineering} of {Photovoltaic} {Devices}
  {XII}}}},\ Vol.\ \bibinfo {volume} {12416}\ (\bibinfo  {publisher} {SPIE},\
  \bibinfo {year} {2023})\ pp.\ \bibinfo {pages} {16--19}\BibitemShut {NoStop}%
\bibitem [{\citenamefont {Ferry}(1974)}]{ferry_decay_1974}%
  \BibitemOpen
  \bibfield  {author} {\bibinfo {author} {\bibfnamefont {D.~K.}\ \bibnamefont
  {Ferry}},\ }\bibfield  {title} {\enquote {\bibinfo {title} {Decay of
  polar-optical phonons in semiconductors},}\ }\href {\doibase
  10.1103/PhysRevB.9.4277} {\bibfield  {journal} {\bibinfo  {journal} {Physical
  Review B}\ }\textbf {\bibinfo {volume} {9}},\ \bibinfo {pages} {4277--4280}
  (\bibinfo {year} {1974})},\ \bibinfo {note} {publisher: American Physical
  Society}\BibitemShut {NoStop}%
\bibitem [{\citenamefont {Ono}(2017)}]{ono_nonequilibrium_2017}%
  \BibitemOpen
  \bibfield  {author} {\bibinfo {author} {\bibfnamefont {S.}~\bibnamefont
  {Ono}},\ }\bibfield  {title} {\enquote {\bibinfo {title} {Nonequilibrium
  phonon dynamics beyond the quasiequilibrium approach},}\ }\href {\doibase
  10.1103/PhysRevB.96.024301} {\bibfield  {journal} {\bibinfo  {journal}
  {Physical Review B}\ }\textbf {\bibinfo {volume} {96}},\ \bibinfo {pages}
  {024301} (\bibinfo {year} {2017})},\ \bibinfo {note} {publisher: American
  Physical Society}\BibitemShut {NoStop}%
\bibitem [{\citenamefont {Bhatt}, \citenamefont {Kim},\ and\ \citenamefont
  {Stroscio}(1994)}]{bhatt_theoretical_1994}%
  \BibitemOpen
  \bibfield  {author} {\bibinfo {author} {\bibfnamefont {A.~R.}\ \bibnamefont
  {Bhatt}}, \bibinfo {author} {\bibfnamefont {K.~W.}\ \bibnamefont {Kim}}, \
  and\ \bibinfo {author} {\bibfnamefont {M.~A.}\ \bibnamefont {Stroscio}},\
  }\bibfield  {title} {{\selectlanguage {english}\enquote {\bibinfo {title}
  {Theoretical calculation of longitudinal-optical-phonon lifetime in
  {GaAs}},}\ }}\href {\doibase 10.1063/1.358498} {\bibfield  {journal}
  {\bibinfo  {journal} {Journal of Applied Physics}\ }\textbf {\bibinfo
  {volume} {76}},\ \bibinfo {pages} {3905--3907} (\bibinfo {year}
  {1994})}\BibitemShut {NoStop}%
\bibitem [{\citenamefont {Tsai}(2018)}]{tsai_effects_2018}%
  \BibitemOpen
  \bibfield  {author} {\bibinfo {author} {\bibfnamefont {C.-Y.}\ \bibnamefont
  {Tsai}},\ }\bibfield  {title} {{\selectlanguage {english}\enquote {\bibinfo
  {title} {Effects of longitudinal optical phonon lifetimes on hot-carrier
  solar cells: a theoretical study},}\ }}\href {\doibase
  10.1088/2053-1591/aaddd2} {\bibfield  {journal} {\bibinfo  {journal}
  {Materials Research Express}\ }\textbf {\bibinfo {volume} {5}},\ \bibinfo
  {pages} {116206} (\bibinfo {year} {2018})},\ \bibinfo {note} {publisher: IOP
  Publishing}\BibitemShut {NoStop}%
\bibitem [{\citenamefont {Hirst}\ \emph {et~al.}(2014)\citenamefont {Hirst},
  \citenamefont {Fujii}, \citenamefont {Wang}, \citenamefont {Sugiyama},\ and\
  \citenamefont {Ekins-Daukes}}]{hirst_hot_2014}%
  \BibitemOpen
  \bibfield  {author} {\bibinfo {author} {\bibfnamefont {L.~C.}\ \bibnamefont
  {Hirst}}, \bibinfo {author} {\bibfnamefont {H.}~\bibnamefont {Fujii}},
  \bibinfo {author} {\bibfnamefont {Y.}~\bibnamefont {Wang}}, \bibinfo {author}
  {\bibfnamefont {M.}~\bibnamefont {Sugiyama}}, \ and\ \bibinfo {author}
  {\bibfnamefont {N.~J.}\ \bibnamefont {Ekins-Daukes}},\ }\bibfield  {title}
  {\enquote {\bibinfo {title} {Hot {Carriers} in {Quantum} {Wells} for
  {Photovoltaic} {Efficiency} {Enhancement}},}\ }\href {\doibase
  10.1109/JPHOTOV.2013.2289321} {\bibfield  {journal} {\bibinfo  {journal}
  {IEEE Journal of Photovoltaics}\ }\textbf {\bibinfo {volume} {4}},\ \bibinfo
  {pages} {244--252} (\bibinfo {year} {2014})}\BibitemShut {NoStop}%
\bibitem [{\citenamefont {Ferry}(2021)}]{ferry_non-equilibrium_2021}%
  \BibitemOpen
  \bibfield  {author} {\bibinfo {author} {\bibfnamefont {D.~K.}\ \bibnamefont
  {Ferry}},\ }\bibfield  {title} {\enquote {\bibinfo {title} {Non-equilibrium
  longitudinal optical phonons and their lifetimes},}\ }\href {\doibase
  10.1063/5.0044374} {\bibfield  {journal} {\bibinfo  {journal} {Applied
  Physics Reviews}\ }\textbf {\bibinfo {volume} {8}},\ \bibinfo {pages}
  {021324} (\bibinfo {year} {2021})}\BibitemShut {NoStop}%
\bibitem [{\citenamefont {Aliberti}\ \emph {et~al.}(2010)\citenamefont
  {Aliberti}, \citenamefont {Feng}, \citenamefont {Takeda}, \citenamefont
  {Shrestha}, \citenamefont {Green},\ and\ \citenamefont
  {Conibeer}}]{aliberti_investigation_2010}%
  \BibitemOpen
  \bibfield  {author} {\bibinfo {author} {\bibfnamefont {P.}~\bibnamefont
  {Aliberti}}, \bibinfo {author} {\bibfnamefont {Y.}~\bibnamefont {Feng}},
  \bibinfo {author} {\bibfnamefont {Y.}~\bibnamefont {Takeda}}, \bibinfo
  {author} {\bibfnamefont {S.~K.}\ \bibnamefont {Shrestha}}, \bibinfo {author}
  {\bibfnamefont {M.~A.}\ \bibnamefont {Green}}, \ and\ \bibinfo {author}
  {\bibfnamefont {G.}~\bibnamefont {Conibeer}},\ }\bibfield  {title} {\enquote
  {\bibinfo {title} {Investigation of theoretical efficiency limit of hot
  carriers solar cells with a bulk indium nitride absorber},}\ }\href {\doibase
  10.1063/1.3494047} {\bibfield  {journal} {\bibinfo  {journal} {Journal of
  Applied Physics}\ }\textbf {\bibinfo {volume} {108}},\ \bibinfo {pages}
  {094507} (\bibinfo {year} {2010})}\BibitemShut {NoStop}%
\bibitem [{\citenamefont {Conibeer}\ \emph {et~al.}(2008)\citenamefont
  {Conibeer}, \citenamefont {König}, \citenamefont {Green},\ and\
  \citenamefont {Guillemoles}}]{conibeer_slowing_2008}%
  \BibitemOpen
  \bibfield  {author} {\bibinfo {author} {\bibfnamefont {G.~J.}\ \bibnamefont
  {Conibeer}}, \bibinfo {author} {\bibfnamefont {D.}~\bibnamefont {König}},
  \bibinfo {author} {\bibfnamefont {M.~A.}\ \bibnamefont {Green}}, \ and\
  \bibinfo {author} {\bibfnamefont {J.~F.}\ \bibnamefont {Guillemoles}},\
  }\bibfield  {title} {\enquote {\bibinfo {title} {Slowing of carrier cooling
  in hot carrier solar cells},}\ }\href {\doibase 10.1016/j.tsf.2007.12.102}
  {\bibfield  {journal} {\bibinfo  {journal} {Thin Solid Films}\ }\bibinfo
  {series} {Proceedings on {Advanced} {Materials} and {Concepts} for
  {Photovoltaics} {EMRS} 2007 {Conference}, {Strasbourg}, {France}},\ \textbf
  {\bibinfo {volume} {516}},\ \bibinfo {pages} {6948--6953} (\bibinfo {year}
  {2008})}\BibitemShut {NoStop}%
\bibitem [{\citenamefont {Garg}\ and\ \citenamefont
  {Sellers}(2020)}]{garg_phonon_2020}%
  \BibitemOpen
  \bibfield  {author} {\bibinfo {author} {\bibfnamefont {J.}~\bibnamefont
  {Garg}}\ and\ \bibinfo {author} {\bibfnamefont {I.~R.}\ \bibnamefont
  {Sellers}},\ }\bibfield  {title} {{\selectlanguage {english}\enquote
  {\bibinfo {title} {Phonon linewidths in {InAs}/{AlSb} superlattices derived
  from first-principles—application towards quantum well hot carrier solar
  cells},}\ }}\href {\doibase 10.1088/1361-6641/ab73f0} {\bibfield  {journal}
  {\bibinfo  {journal} {Semiconductor Science and Technology}\ }\textbf
  {\bibinfo {volume} {35}},\ \bibinfo {pages} {044001} (\bibinfo {year}
  {2020})},\ \bibinfo {note} {publisher: IOP Publishing}\BibitemShut {NoStop}%
\end{thebibliography}%

\end{document}